# Open SYCL on heterogeneous GPU systems: A case of study

Rocío Carratalá-Sáez[a,*], Francisco J. Andújar[a], Yuri Torres[a], Arturo Gonzalez-Escribano[a], Diego R. Llanos[a]

[a]*Universidad de Valladolid, Dpto. de Informática, Paseo de Belén 15, Valladolid, 47011, Spain*

## Abstract

Computational platforms for high-performance scientific applications are becoming more heterogenous, including hardware accelerators such as multiple GPUs. Applications in a wide variety of scientific fields require an efficient and careful management of the computational resources of this type of hardware to obtain the best possible performance. However, there are currently different GPU vendors, architectures and families that can be found in heterogeneous clusters or machines. Programming with the vendor provided languages or frameworks, and optimizing for specific devices, may become cumbersome and compromise portability to other systems. To overcome this problem, several proposals for high-level heterogeneous programming have appeared, trying to reduce the development effort and increase functional and performance portability, specifically when using GPU hardware accelerators.

This paper evaluates the SYCL programming model, using the Open SYCL compiler, from two different perspectives: The performance it offers when dealing with single or multiple GPU devices from the same or different vendors, and the development effort required to implement the code. We use as case of study the Finite Time Lyapunov Exponent calculation over two real-world scenarios and compare the performance and the development effort of its Open SYCL-based version against the equivalent versions that use CUDA or HIP.

Based on the experimental results, we observe that the use of SYCL does not lead to a remarkable overhead in terms of the GPU kernels execution time. In general terms, the Open SYCL development effort for the host code is lower than that observed with CUDA or HIP. Moreover, the SYCL version can take advantage of both CUDA and AMD GPU devices simultaneously much easier than directly using the vendor-specific programming solutions.

## 1. Introduction

The complexity of the scientific applications follows an increasing trend motivated by the society needs. Arising from many fields, the computational applications require as much computational power as possible to efficiently contribute to the scientific, commercial and social progress. To accomplish this, high performance computing (HPC) is vital. HPC relies on the efficient usage of the diversity of resources available in modern computational systems, that are becoming more and more heterogeneous. This includes not only traditional multicore systems, but also the exploitation of devices such as Graphic Processing Units (GPU), among others. In the particular case of GPUs, it has been proved that they offer great computational capabilities that can accelerate many computations by several orders of magnitude.

To take advantage of all the available hardware in a heterogeneous system, the first approach is usually to manually develop a specific solution for that particular hardware, using the vendor toolchains or parallel programming models. For example, CUDA [1] for NVIDIA GPUs, or HIP [2] for AMD GPUs. These tools and models have demonstrated great capabilities and a great versatility to obtain the best possible performance for those devices, thanks to efficiently managing the hardware resources. Nevertheless, experts that do not belong to the HPC field, such as other engineers, physicists or mathematicians, have to deal with a non-negligible learning curve to take advantage of all the capabilities of these programming models. Moreover, using vendor specific tools the resulting applications are often not easily portable to alternative vendor devices, and additional programming efforts are needed to use different hardware.

In recent years, different approaches with an increasing level of abstraction have been presented for designing applications that can leverage the resources in heterogeneous systems with improved portability. OpenCL [3] is a good example of approaches that introduce a first layer of abstractions for dealing with heterogeneous devices. It is an extension of the C/C++ programming language, capable of generating and running applications on multiprocessors, FPGAs and GPUs of different vendors. However, OpenCL requires even a higher development effort than, for example, the use of vendor-specific programming models for GPUs, such as CUDA or HIP. Moreover, OpenCL requires to explicitly manage the data transfers and synchronization using a low-level event model, further increasing the development effort if the programmer wants to perform asynchronous operations in order to overlap kernel executions and data transfers. For this reason, learning and using OpenCL is cumbersome for those who are not HPC experts, but want to maximize their intensive-computation applications by exploiting the available resources in different heterogeneous environments.

[*]Corresponding author
  *Email addresses:* rocio@infor.uva.es (Rocío Carratalá-Sáez),
fandujarm@infor.uva.es (Francisco J. Andújar),
yuri.torres@infor.uva.es (Yuri Torres), arturo@infor.uva.es
(Arturo Gonzalez-Escribano), diego@infor.uva.es (Diego R. Llanos)

In contrast, there are other proposals for higher-level heterogeneous programming such as SYCL [4], OpenMP [5], Kokkos [6], Raja [7], or other more academic approaches such as dOCAL [8] or CtrlEvents [9] that pursue a common objective: Offering higher-level abstractions that simplify and unify the programming of different computational resources in a transparent and effortless way. While OpenMP is widely available in most modern compilers and the other alternatives previously cited have specific advantages, SYCL is becoming more and more popular as the available compiler implementations are becoming more mature, complete, robust and efficient (see e.g., Open SYCL [10], or Intel oneAPI DPC++ [11]). SYCL advocates a single-code approach, with automatic data-dependence analysis and data movements across memory hierarchies, which are easy to understand and to program by non-experts in low-level programming of heterogeneous devices. The SYCL community is striving to make it the baseline for functional and performance portability. As we discuss in Section 3, several works compare the efficiency and portability between SYCL and other heterogeneous programming models for specific applications and platforms. Currently, it is highly relevant to investigate the efficiency and portability offered by the new SYCL implementations for real-world applications.

In this paper, we evaluate the current Open SYCL implementation from two different perspectives: The performance it offers when dealing with single or multiple GPU devices, from the same or different vendors, and the development effort required to implement the code. We compare the performance and the code with baselines programmed directly using CUDA or HIP technologies for NVIDIA and AMD GPUs, both isolated or in combination. We use as case of study a real-world application. With this comparison, we pursue to shed some light on the advantages and limitations of using the recent improvements introduced for this high-level programming model, in comparison with using the traditional vendor provided tools.

We have chosen as case of study the UVaFTLE [12] application, which computes the Finite Time Lyapunov Exponent (FTLE), as the mean to explore this development effort and performance evaluation. On the one hand, this application is formed by two kernels that are conceptually very different: One deals with larger data sets and memory accesses, while the other one focuses on solving a collection of linear algebra operations. This difference lets us explore whether the key aspects of most of the scientific applications (memory accesses and computations) are better addressed by native (vendor-provided) tools than by Open SYCL. On the other hand, we have not found any work in the literature that offers a recent and portable version of the FTLE solution, so we also provide the community with a novel portable and improved FTLE implementation, based on our previous work [12].

The main contributions of this work are:

- We present a portable version of the UVaFTLE application using Open SYCL, with support to target multiple GPU devices simultaneously, even from different vendors.

- We present new baseline implementations of the UVaFTLE application. The first one uses CUDA. It improves a previous version [12] with the use of pinned memory for faster memory transfers, a more intense use of registers to minimize global memory accesses, and a new kernel to implement the data preprocessing stage in GPU. The second baseline is a port of the same program using HIP, to target AMD GPU devices. Both versions support multi-GPU of the specific vendor.

- We conduct an in-depth evaluation of the performance, in terms of execution time, offered by both the baseline implementations of the FTLE computation (based on CUDA and HIP) and the new Open SYCL version.

- We compare the development effort required to implement the CUDA and HIP baselines with the Open SYCL version, in terms of several classical development-effort metrics.

- This work contributes to open science. All our implementations are fully open-source and available by accessing the GitHub repository [13].

The rest of the paper is structured as follows: In Section 2 we provide a revision of the different SYCL implementations and the mathematical background of the FTLE; in Section 3 we summarize the main existing works that use SYCL in their implementations, as well as those related to the FTLE computation; in Section 4 we describe the FTLE computation algorithm and our implementations, covering how do we leverage CUDA, HIP and Open SYCL; in Section 5 we present an in-depth evaluation of the different implementations' performance (in terms of execution time); in Section 6 we analyze the development effort associated to each implementation; and in Section 7 we summarize the main conclusions derived from this work and finalize by mentioning the future work lines.

## 2. Background

In this section, we summarize the state of the art of SYCL, describing its different implementations, as well as the main features of each of them. After that, we describe the case of study we utilize in this work: Finite Time Lyapunov Exponent (FTLE).

### 2.1. Heterogeneous computing and SYCL

In 2014, the Khronos Group presented SYCL [4], a standard model for cross-platform programming, with the purpose of achieving both code and performance portability, and lowering the development effort. SYCL organizes the kernels using a task graph implicitly constructed by the SYCL runtime. This also allows to implicitly manage the dependencies between the kernels and the data communications, although the developer can still manage them explicitly.

The SYCL ecosystem has several SYCL implementations, being the most important compilers Codeplay's ComputeCPP [14], Intel's OneAPI [11], TriSYCL [15], and Open SYCL [16, 17] (formerly known as HipSYCL). However, these implementations rely on different compiler back-ends for different types of devices, and, therefore, each one has support for different hardware. TriSYCL only supports CPUs through OpenMP or TBB, and Xilinx FPGAs. ComputeCPP supports CPUs, NVIDIA GPUs through OpenCL+PTX, and Intel CPUs, Intel GPUs and AMD GPUs through OpenCL+SPIR-V, although the latest AMD GPU drivers do not support SPIR-V. Regarding OneAPI, it only supports Intel hardware (CPUs, GPUs, and FPGAs), although there is a project to support NVIDIA devices using an alternative CUDA backend through LLVM [18]. However, this backend is not compatible with the rest of Intel hardware. Finally, Open SYCL supports CPUs, NVIDIA GPUs, AMD GPUs, and Intel GPUs through OpenMP, CUDA, HIP/ROCm, and Level Zero, respectively.

Moreover, there are multiple ways of implementing the SYCL compiler. According to the SYCL specification, there are three different choices:

- **Library only-implementation**: It is possible to implement SYCL as a pure C++ library. For example, this approach is available in TriSYCL and Open SYCL to support host CPU code, and to, besides, target NVIDIA GPUs in Open SYCL.

- **Single-source, single-compiler pass** (SSCP): The host and the device binary is generated from a unique SYCL code and a unique compiler invocation. Open SYCL has recently presented the first version of a SSCP SYCL compiler [10].

- **Single-source, multiple-compiler passes** (SMCP): The host and the device binaries are generated from a unique SYCL code, but it is necessary to compile the device code several times (once per specific SYCL device), generating different device images inside the application binary. This approach is the most frequently implemented one, but it requires a higher compilation time.

As it can be seen, Open SYCL is one of the most complete SYCL compilers. It allows compiling SYCL codes on the main currently available GPUs (AMD, NVIDIA, and Intel) generating a unique application binary, which is not possible with OneAPI or TriSYCL. Regarding ComputeCPP, it only has support for AMD GPUs in older models that support SPIR-V. Moreover, Codeplay announced that there will no longer be support for ComputeCPP after September 2023 [19]. For these reasons, the Open SYCL compiler has been chosen for conducting this study.

## 2.2. Case of study: FTLE

Fluid dynamics is a widely explored field. In particular, the fluid particle trajectories in phase space, often referred to as *Lagrangian*, is of great interest. More specifically, calculating the *Lagrangian Coherent Structures (LCS)* [20] is key for several disciplines, such as cardiovascular engineering [21], aerodynamics [22], and geophysical fluid dynamics [23].

The fluid particle trajectories are defined as solutions of

$$\dot{\vec{x}} = \vec{v}(\vec{x}, t),$$

where the right-hand side is the velocity field of the fluid, in absence of molecular diffusion. Solving this system of equations allows calculating the LCS. The main interest on computing the LCS is the fact that they let a better understanding of the flow phenomena, since they can be broadly interpreted as *transport barriers* in the flow.

From the computational point of view, the extraction of LCS consists of two main steps: The flowmap computation and the resolution of the FTLE. We will focus on the second step, which is mathematically defined as

$$\Lambda_{t_0}^{t_1}(\vec{x}_0) = \frac{1}{t_1 - t_0} \log \sqrt{\lambda_n(\vec{x}_0)}$$

where $\lambda_n$ is the maximum eigenvalue of the Cauchy-Green strain tensor $C$, defined as follows

$$C(\vec{x}_0) = \left[\nabla F_{t_0}^{t_1}(\vec{x}_0)\right]^T \nabla F_{t_0}^{t_1}(\vec{x}_0)$$

being $F$ the flowmap [22].

The FTLE is a scalar field that works as an objective diagnostic for LCS: A first-order approach to assess the stability of material surfaces in the flow under study, by detecting material surfaces along which infinitesimal deformation is larger or smaller than off these surfaces [20]. Although more reliable mathematical methods have been developed for the explicit identification of LCS, the FTLE remains the most used metric in the field for LCS identification.

From the computational point of view, it is important to highlight that the FTLE computation is applied to each particle of the flow independently of the other particles. Thus, it represents an embarrassingly-parallel problem [24]. We have already described, explored, and evaluated the FTLE computation in a previous work [12], where we presented UVaFTLE, a tool that incorporates a CUDA-based kernel to use multiple NVIDIA GPUs in the FTLE computation.

## 3. Related work

In this section, we briefly describe the main existing contributions that leverage SYCL and study their functional and/or performance portability, as well as the works that focus on the FTLE computation and their limitations.

*3.1. SYCL portability*

Due to the growing interest in heterogeneous computing and SYCL, there are several works using this standard and studying its portability. Some of these works are focused on code migration to SYCL from other languages like CUDA [25, 26, 27], OpenCL [28, 29], or OpenMP [30], comparing the performance of both versions. Other papers present SYCL libraries to speed up and make portable other scientific works, such as machine learning [31], or neural network [32] algorithms; or present a SYCL hand-tuned version of a specific algorithm, comparing it with the state-of-the-art algorithm [33].

Other works are focused on the performance evaluation of SYCL compilers. In [34], the authors made a comparative study of OpenCL, OpenMP and TriSYCL in multiprocessors. However, TriSYCL currently does not have support for GPUs. In [35], a comparison using several benchmarks and the Intel LLVM-SYCL compiler against CUDA using Tesla V-100 is presented. However, AMD architecture are not studied. Other works compare several SYCL compilers [36, 37, 38] against multiple AMD and NVIDIA GPUs models.

To the best of our knowledge, none of the existing works explore the possibilities offered by SYCL using multiple GPUs of both NVIDIA and AMD architectures simultaneously, also analyzing the implications on the development effort of coding in SYCL instead of CUDA or HIP.

*3.2. FTLE computation*

In the literature, there are previous works that offer optimizations in the context of the FTLE computation. Some [39, 40, 41, 42, 43] focus on speeding up the calculations of the FTLE by applying some optimization techniques such as reducing I/O, optimizing the use of the memory hierarchy, or using multiple CPUs. Other authors [44, 45, 46, 47, 48, 49] focus on exploiting GPU devices to accelerate the FTLE computation. Another study proposes the use of an Accelerated Processing Unit (APU) to speed up the computation of FTLEs [50].

As we described in our previous work [12], the main problems of the existing proposals that leverage GPU devices to compute the FTLE are that most of them are old and based on outdated tools which are not capable of tackling nowadays devices. Besides it, in general, multi-GPU scheme is not supported. Moreover, neither an in-depth description of the GPU implementation or the source code are provided. For these reasons, in our previous work we offered a competitive, open-source implementation of the FTLE computation (named UVaFTLE) equipped with a CUDA kernel capable of simultaneously using multiple NVIDIA GPU devices.

To the best of our knowledge, in the existing literature, there is a lack of updated proposals of the FTLE computation that tackle heterogeneous environments provided with GPU devices from different vendors. To fill this gap, in this work we redesign UVaFTLE to use Open SYCL in such a way that it can leverage any GPU device, regardless of its vendor. For completeness, we also present a novel UVaFTLE implementation that uses HIP instead of CUDA to tackle AMD GPU devices. Moreover, we evaluate the Open SYCL performance compared to that offered by the implementations based on HIP or CUDA.

**4. Our implementations**

In this section, we describe the FTLE algorithm, next we identify the regions of code suitable to be executed in GPUs, afterward we present the native (CUDA and HIP) and the Open SYCL implementations of the GPU kernels, and, finally, we illustrate how to target multiple GPUs using Open SYCL. Note that the full code of all versions is available in the UVaFTLE repository [13].

*4.1. FTLE algorithm*

Provided the information of the mesh that defines the flow to study (namely the dimension, time instant when the FTLE will be computed, the mesh points coordinates and faces information, and the flowmap), the process of computing the FTLE (described in Algorithm 1) consists of the following steps performed over each point in the mesh:

1. Compute the gradients of the flowmap (see Algorithm 2). Note that calculating the gradients is done based on the Green Gauss theorem [51].
2. Generate the tensors from the gradients and perform the matrix-matrix product of the previously generated tensors by their transposes (see Algorithm 2).
3. Compute the maximum eigenvector of each resulting matrix (see Algorithm 3). Note that, as we are computing the eigenvalues of matrices of size 2x2 (2D) or 3x3 (3D), which in practice means respectively solving a 2nd and 3rd degree equation, we have directly implemented this computation, instead of calling mathematical libraries that perform this computation for generic matrices of any size.
4. Calculate the logarithm of the square matrix of the maximum eigenvalue and divide the result by the time instant to evaluate.

Note that we are only presenting here the algorithms for the 2D case because the 3D case is straightforward.

In addition to the algorithms already described, it is also important to remark those utilized in lines 5 and 6 in Algorithm 1: *create_nFacesPerPoint_vector* (see Algorithm 4) and *create_FacesPerPoint_vector* (see Algorithm 5). Although they are part of the preprocessing and not the FTLE computation itself, they are needed to create the data structures called nFpP and FpP, that respectively contain the number of faces to which each mesh point belongs, and the corresponding faces identifiers. These data structures serve to accelerate the process of computing the FTLE, because they establish the relationship between the different mesh points and faces, meaning that this is analyzed only once at the beginning of the code, instead of each time the Green Gauss algorithm is called.

*4.2. GPU kernels identification*

The cost of computing the FTLE algorithm described in the previous section relies on two main procedures: The *create_facesPerPoint_vector* function and the linear algebra operations performed for each mesh point in each iteration of the *for* loop in line 7 of the Algorithm 1. As a consequence, this is what is worth it to be computed in the GPU; in other words, these are the two GPU kernels to build in order to accelerate the FTLE computation:

- **Preprocessing**: This kernel directly implements the *create_facesPerPoint_vector* function (see Algorithm 5).

- **FTLE**: This kernel was already described in our previous work [12]; we presented a single CUDA based kernel to compute everything described in Algorithms 2 and 3 (or their corresponding 3D versions), which means using the GPU device to compute lines 9-10 (2D case) or 12-13 (3D case) of the Algorithm 1. Note that this kernel has two variants: 2D and 3D.

In the following sections, we present details regarding how to implement these kernels using CUDA or HIP (namely native implementations) and Open SYCL.

---

**Algorithm 1** FTLE

**Require:** $nDim, t\_eval, coords\_file, faces\_file, flowmap\_file$
1: $nVpF = (nDim == 2)\ ?\ 3 : 4$ ▷ Triangles or tetrahedrons
2: $\{nPoints, coords\}$ = read_coordinates(coords_file)
3: $\{nFaces, faces\}$ = read_faces(faces_file)
4: $flow$ = read_flowmap(flowmap_file)
5: $nFpP$ = create_nFacesPerPoint_vector(nPoints, nFaces, nVpF, faces)
6: $FpP$ = create_FacesPerPoint_vector(nPoints, nFaces, nVpF, faces, nFpP)
7: **for** $ip$ in range(nPoints) **do**
8:    **if** $nDim == 2$ **then**
9:       $g1, g2$ = 2D_grad_tens (ip, nVpF, coords, flow, faces, nFpP, FpP)
10:      $max\_eigen$ = max_eigenvalue_2D([g1, g2])
11:    **else**
12:       $g1, g2, g3$ = 3D_grad_tens (ip, nVpF, coords, flow, faces, nFpP, FpP)
13:      $max\_eigen$ = max_eigenvalue_3D([g1, g2, g3])
14:    **end if**
15:    $result[ip] = log(sqrt(max\_eigen))/t\_eval$
16: **end for**
17: **return** $result[\ ]$

---

**Algorithm 2** 2D_grad_tens

**Require:** $ip, nP, nVpF, coords[\ ], flow, faces[\ ], nFpP[\ ], FpP[\ ]$
1: $nFaces = (ip == 0)\ ?\ nFpP[ip] : nFpP[ip] - nFpP[ip-1]$
2: $left, right, below, above$ = GreenGauss(nFaces, FpP, nFpP, nVpF, coords)
  ▷ This provides the indices of the left, right, below, above closest points
3: $dx = coords[right \cdot nDim] - coords[left \cdot nDim]$
4: $dy = coords[above \cdot nDim + 1] - coords[below \cdot nDim + 1]$
5: $gra1[0] = (flow[right \cdot nDim] - flow[left \cdot nDim])/dx$
6: $gra1[1] = (flow[right \cdot nDim + 1] - flow[left \cdot nDim + 1])/dx$
7: $gra2[0] = (flow[above \cdot nDim] - flow[below \cdot nDim])/dy$
8: $gra2[1] = (flow[above \cdot nDim + 1] - flow[below \cdot nDim + 1])/dy$
9: $ftle\_m[0] = gra1[0] \cdot gra1[0] + gra1[1] \cdot gra1[1]$
10: $ftle\_m[1] = gra1[0] \cdot gra2[0] + gra1[1] \cdot gra2[1]$
11: $ftle\_m[2] = gra2[0] \cdot gra1[0] + gra2[1] \cdot gra1[1]$
12: $ftle\_m[3] = gra2[0] \cdot gra2[0] + gra2[1] \cdot gra2[1]$
13: $gra1[0] = ftle\_m[0]; gra1[1] = ftle\_m[1]$
14: $gra2[0] = ftle\_m[2]; gra2[1] = ftle\_m[3]$
15: $ftle\_m[0] = gra1[0] \cdot gra1[0] + gra1[1] \cdot gra1[1]$
16: $ftle\_m[1] = gra1[0] \cdot gra2[0] + gra1[1] \cdot gra2[1]$
17: $ftle\_m[2] = gra2[0] \cdot gra1[0] + gra2[1] \cdot gra1[1]$
18: $ftle\_m[3] = gra2[0] \cdot gra2[0] + gra2[1] \cdot gra2[1]$
19: **return** $ftle\_m$

---

**Algorithm 3** max_eigenvalue_2D

**Require:** $M$
1: $sq \leftarrow sqrt(M[21] * M[21] + M[10] * M[10] - 2 * (M[10] * M[21]) + 4 * (M[11] * M[20]))$
2: $eig1 \leftarrow (M[21] + M[10] + sq)/2$
3: $eig2 \leftarrow (M[21] + M[10] - sq)/2$
4: **return** $(eig1 > eig2)\ ?\ eig1 : eig$

---

**Algorithm 4** create_nFacesPerPoint_vector

**Require:** $nPoints, nFaces, nVpF, faces[\ ]$
1: **for** $ip$ in range(nPoints) **do**
2:    $nFpP[ip] = 0;$
3: **end for**
4: **for** $iface$ in range(nFaces) **do**
5:    **for** $ipf$ in range(nVpF) **do**
6:       $ip = faces[iface \cdot nVpF + ipf]$
7:       $nFpP[ip] = nFpP[ip] + 1$
8:    **end for**
9: **end for**
10: **for** $ip$ in range(nPoints) **do**
11:    $nFpP[ip] = nFpP[ip] + nFpP[ip-1]$
12: **end for**
13: **return** $nFpP$

---

**Algorithm 5** create_facesPerPoint_vector

**Require:** $nPoints, nFaces, nVpF, faces[\ ], nFpP[\ ]$
1: **for** $ip$ in range(nPoints) **do**
2:    $count = 0$
3:    $iFacesP = (ip == 0)\ ?\ 0 : nFpP[ip-1]$
4:    $nFacesP = (ip == 0)\ ?\ nFpP[ip] : nFpP[ip] - nFpP[ip-1]$
5:    **while** $(iface < nFaces)\ \&\ (count < nFacesP)$ **do**
6:       **for** $ipf$ in range(nVpF) **do**
7:          **if** $faces[iface \cdot nVpF + ipf] == ip$ **then**
8:             $FpP[ifacesP + count] = iface$
9:             $count = count + 1$
10:         **end if**
11:       **end for**
12:    **end while**
13: **end for**
14: **return** $FpP$

---

*4.3. Native implementations*

Three different GPU kernels (*create_facesPerPoint_vector*, *gpu_compute_gradient_2D*, and *gpu_compute_gradient_3D*) have been developed corresponding to the algorithms described in previous sections. The *gpu_compute_gradient_2D* and the *gpu_compute_gradient_3D* kernels are improved versions of the CUDA-based implementation of our previous work, UVaFTLE [12]. Moreover, they have been appropriately ported to HIP in order to tackle AMD GPUs.

All three kernels, regardless of using CUDA or HIP, perform the same two initial operations before starting the algorithm. The first operation corresponds to the calculation of the thread global identifier. Each identifier corresponds to a mesh point. For the code simplicity, we use one-dimensional `threadBlock` and `grid`, making it easier to calculate the global index of each thread, reducing the number of kernel instructions. The following instruction is executed to calculate the thread global identifier:

    *int th_id = blockIdx.x * blockDim.x + threadIdx.x;*

The second operation checks that the number of threads that are launched is not larger than the points contained in the mesh. For that, we insert the following condition wrapping each kernel implementation:

$$if(th\_id < numCoords)\ \{...\}$$

For each kernel, each thread of the GPU grid executes exactly the sequence of steps associated to the FTLE kernel described in Section 4.2. The implementation is currently capable of leveraging all the GPU devices available in a single node, as in our previous work [12]. Thus, we are deploying our multi-GPU executions in a shared-memory environment. We use the OpenMP programming model, instantiating as many threads as GPU devices to distribute the load among them. Particularly, we have designed a static partitioning of the mesh points based on the number of GPU devices that take part of the execution.

In contrast to our previous work, pinned memory has been used to perform the data transfers of the results from the GPU to the host through *cudaHostAlloc* or *hipHostAlloc* primitives. Classical GPU reference manuals, such as [1], indicate that this kind of memory can be used when executions or asynchronous transfers are introduced, thus reducing the latencies in these data transfers.

Furthermore, the GPU community indicates that the best `threadBlock` size is one that maximizes the streaming multiprocessor occupancy, such as 256, 512, and 1024. We have selected 512 as the `threadBlock` size, since it is one of the recommended ones. As this work does not intend to apply any tuning strategies, we have not evaluated additional sizes.

*4.4. Porting UVaFTLE to Open SYCL*

On the basis of the native implementations, the application has been ported to SYCL. Since the full code of UVaFTLE is very large, we will illustrate the changes made in our application using a simpler code. Note that the full SYCL code of the UVaFTLE can be found in our repository [13].

Figure 1 shows the code examples, which launch a simple kernel that, given an array *A* with *n* elements, calculates $A[i] = 2 \times A[i] + 1$ for each element *i*, being $0 \leq i < n$. Figures 1a and 1b depict the CUDA and SYCL code, respectively. The background of both codes has been colored to make it easier for the reader to identify the groups of lines in both codes that have the same functionality. The parts with white background correspond to the host code, and there are no differences between both versions. Also note that the HIP code has not been included in the comparison, since the differences with the CUDA and HIP versions are practically negligible.

In first place, we need to choose the device to execute the code (code with blue background). For these purposes, SYCL employs a *queue*, which is an abstraction to which the kernels that are going to be executed on a single device are submitted. This is performed in line 9 of Figure 1b, where a new queue is created and attached to a GPU device. Note that, through the usage of *gpu_selector*{}, the kernel to be executed can be attached to any GPU in the system (usually the first GPU detected by the SYCL runtime). However, the SYCL API offers methods to attach a GPU from a specific platform, a specific model, etc. For example, Figure 2 shows a function for creating a queue attached to a HIP device, getting at first the list of devices for the HIP platform. Attaching the queue to a CUDA device is also possible, simply comparing the string "CUDA" with the platform name.

In the second part of the code (code with purple background), the native implementation specifies the CUDA `numBlocks` and `grid` sizes. In SYCL, we must specify the range of our arrays (*array_range* in the example) and the range of the thread block (*block_range*). *array_range* will be used later to create the buffer. Both ranges will be necessary to launch the kernel. Therefore, we can create all the necessary ranges to port our application to SYCL. Note that, for the simplicity of the example, we only use 1-dimensional ranges, but we can also specify 2-dimensional or 3-dimensional ranges.

In the third part, the management of the memory hierarchy is shown (code with green background). While in the native implementation we need to manually allocate and free the device memory, as well as to manually manage the data transfers (both synchronous and asynchronous versions) between host and devices, or between devices, the buffer abstraction simplifies the memory management. A buffer provides an abstract view of the memory that is accessible from the host and the devices. The buffers also allow the SYCL runtime to manage the memory transfers transparently to the programmer.

For example, let's suppose three kernels: $K_1$ and $K_2$, that have no data dependencies, and $K_3$, that needs the results of the first two kernels to make its own work. Using the buffer abstraction, the SYCL runtime transparently transfers the host data to the devices running $K_1$ and $K_2$. Since both kernels have no data dependencies, both kernels can run concurrently in different devices. Once the kernels have finished, the SYCL runtime will transfer the necessary data to run $K_3$ in its device, and finally transfer the resulting data to the host global memory.

To declare a buffer (line 16 of Figure 1b) it is necessary to specify, in the C++ template, the number of dimensions and the data type, and indicate in its constructor the host memory to be managed through this buffer, and the buffer range. Therefore, to port UVaFTLE to SYCL, we have created the necessary buffers to manage all the application data. Finally, note that the buffer is created inside a new scope. The host memory will not be updated until the scope ends, and the buffer is destroyed, although the programmer can manually update the host memory inside the scope.

The buffers are not directly accessed by the programmer in the kernels. To read and write buffers, we must create an *accessor* object (line 20 in Figure 1b), specifying the accessed buffer and the access mode (read, write, or read_write). We specify the main code differences that perform the same functionality in CUDA, HIP, and SYCL in Table 1.

Finally, we specify the kernel declaration (code with dark red background) and its launch (code with light red background). In the native implementation, we should declare the kernel as a function (lines 31-38 in Figure 1a) and launch this function inside the host code using a specific syntax (line 20 in Figure 1a). In SYCL, the *submit()* method is used to submit the kernel

```
 1  int main(){
 2      // Host memory declaration
 3      int elements=100000;
 4      int mem_size=sizeof(float) * elements;
 5      float host_array[elements];
 6      // Host memory initialization
 7      [...]
 8      // Set the device
 9      cudaSetDevice(0);
10      // Declare block and grid
11      dim3 block(512);
12      int numBlocks = (int) (ceil((double)elements/
            block.x)+1);
13      dim3 grid(numBlocks);
14      // Device memory declaration
15      cudaMalloc(dev_array, mem_size);
16      // Copy host memory to device
17      cudaMemcpy(dev_array, host_array, mem_size,
            cudaMemcpyHostToDevice);
18      // Kernel launch
19      my_kernel<<<grid, block, 0, cudaStreamDefault
            >>>(dev_array, elements);
20      // Asynchronous copy from device to host
21      cudaMemcpyAsync (host_array, dev_array,
            mem_size, cudaMemcpyDeviceToHost,
            cudaStreamDefault);
22      // Synchronization
23      cudaDeviceSynchronize();
24      // Free device memory
25      cudaFree(dev_array);
26      // Use the results of the kernel in the host
27      [...]
28      return 0;
29  }
30  // Kernel declaration
31  __global__ void my_kernel (float* array, int
        elements){
32      // Get the thread global identifier
33      int gpu_id = blockIdx.x*blockDim.x +
            threadIdx.x;
34      // Kernel computation
35      if(gpu_id < elements)
36          array[gpu_id] = array[gpu_id]* 2 +1;
37  (\breaklist{31}{kernel}
38  }
```

(a) CUDA code

```
 1  using namespace cl::sycl;
 2  int main(){
 3      // Host memory declaration
 4      int elements=100000;
 5      float host_array[elements];
 6      // Host memory initialization
 7      [...]
 8      // Set the execution queue selecting a GPU
 9      queue my_queue(gpu_selector{});
10      // Range declaration
11      range<1> array_range{static_cast(elements)};
12      range<1> block_range{static_cast(512)};
13      // Enters in a new scope
14      {
15      // Buffer declaration
16      buffer<float,1> dev_buf(host_array,
            array_range);
17      // Submit the kernel to the queue
18      my_queue.submit([&](handler &my_handler) {
19      // Create the accessor to use the device
            array
20      auto array = dev_buff.get_access<
            access_mode::read_write>(my_handler);
21      // Execute the kernel with a parallel for
22      my_handler.parallel_for<class my_kernel>(
            nd_range<1>(array_range,block_range),
            [=](nd_item<1> i){
23          // Get the thread global identifier
24          int gpu_id = i.get_global_id(0);
25          // Kernel computation
26          if(gpu_id < elements)
27              array[gpu_id]=array[gpu_id]*2+1;
28      }); // end parallel for
29      }); // end submit
30      // Finish the scope to update host memory
31      }
32      // Use the results of the kernel in the host
33      [...]
34      return 0;
35  }
```

(b) SYCL code

Figure 1: Comparison between the CUDA (a) and SYCL (b) kernels implementation. Note that the lines with the same colors share purpose in both codes.

```
1  queue getHIPqueue(){
2      auto platform = platform::get_platforms();
3      for (int p=0; p < platform.size(); p++){
4          if(!platform[p].get_info<info::platform::
               name>().compare("HIP")){
5              auto devs= platform[p].get_devices();
6              return queue(devs[0]);
7          }
8      }
9  }
```

Figure 2: Example of a function for getting a SYCL queue attached to a HIP device.

using the desired queue (line 18 in Figure 1b). Using lambda functions, we should specify the necessary accessors to manage the desired buffer, defining the kernel using another lambda function. In the example, a *parallel_for* and *nd_range* kernel (lines 22-28 in Figure 1b)) are employed to perform the same work as the CUDA kernel , i.e., to launch a kernel with *elements* threads organized in blocks of 512 threads. Note that the kernel code is the same in both versions. If we appropriately name the accessor objects, it is not necessary to make changes in our code kernel. The only difference between both codes is how to obtain the index to access the data, as can be seen in line 24 of Figure 1b. Note that, since the main purpose is not to describe the SYCL API, we will not go into more detail about the declaration of lambda functions. For further information, please consult the reference guide [4].

Summarizing, the steps to port UVaFTLE to Open SYCL have been the following:

| **Action** | **Language** | **Function** |
|---|---|---|
| Allocate device memory | CUDA | cudaMalloc(dev_array, mem_size) |
| | HIP | hipMalloc(dev_array, mem_size) |
| | SYCL | ::buffer dev_buf(host_array, range<1>{static_cast(num_elements)}) |
| Copy from host to device | CUDA | cudaMemcpy(dev_array, host_array, mem_size, cudaMemcpyHostToDevice) |
| | HIP | hipMemcpy(dev_array, host_array, mem_size, hipMemcpyHostToDevice) |
| | SYCL | Implicitly done by SYCL runtime when dev_buf is used in a device kernel |
| Access to device memory inside the kernel | CUDA | Declare the array in the kernel prototype and |
| | HIP | include the device array in the kernel invocation |
| | SYCL | Create an accessor in kernel submit<br>auto array = dev_buff.get_access<access::mode::read_write>(my_handler)<br>Use accessor in kernel code |
| Asynchronous copy from device to host | CUDA | cudaMemcpyAsync(host_array, dev_array, mem_size, cudaMemcpyDeviceToHost, cudaStream) |
| | HIP | hipMemcpyAsync(host_array, dev_array, mem_size, hipMemcpyDeviceToHost, hipStream) |
| | SYCL | Implicitly done by SYCL runtime when the scope of dev_buf ends |
| Synchronization to ensure the host memory is updated | CUDA | cudaDeviceSynchronize() |
| | HIP | hipDeviceSynchronize() |
| | SYCL | Implicitly done by SYCL runtime when the scope of dev_buf ends |
| Free device memory | CUDA | cudaFree(dev_array) |
| | HIP | hipFree(dev_array) |
| | SYCL | Implicitly done by SYCL runtime when the scope of dev_buf ends |

Table 1: Memory management in CUDA, HIP and SYCL

1. Copy the original host code (mainly, the declaration and initialization of the host memory, as well as storing the final results of the application).
2. Create a *queue* attached to the desired GPU device.
3. Start a new scope and define the *buffers* to manage the application data.
4. Submit the prepossessing kernel to the *queue*.
   (a) Create the *accessors* with the appropriate names to avoid rewriting the kernel code.
   (b) Launch the kernel using an *nd-range parallel for*.
   (c) Copy the kernel code, changing the index calculation to SYCL syntax.
5. Submit the FTLE kernel to the *queue*, repeating the sub-steps of step 4.
6. End the scope to update the host memory, allowing the host to get the final results.

### 4.5. Targeting multiple GPUs and vendors with Open SYCL

At this point, UVaFTLE has been ported to Open SYCL and can be executed in NVIDIA and AMD GPUs. However, the application still does not have support for multi-GPU execution. From now on, we will use the term "*sub-kernel*" to refer to one part of a single kernel distributed across different devices, while the term "*kernel*" will refer to the execution of all the parts of the kernel. The native application uses OpenMP to instance multiple threads, and each thread performs a part of the computational work or sub-kernel, using a different GPU device, as explained in Section 4.3. However, this solution is not possible in our case, since SYCL kernels can not be used inside OpenMP target regions [52].

Fortunately, we can do the same job instantiating as many SYCL queues as devices we need, and attaching each queue to a different device. Moreover, the queue abstraction allows us to use GPUs from different architectures, such as NVIDIA and AMD. For example, the function shown in Figure 2 could be easily modified to get a vector of queues with all the AMD GPUs attached to the current node, and Figure 3 shows a function that returns a queue vector to use all the GPUs in the node, regardless of their vendor or architecture. If the program was compiled targeting all the GPUs on the system, the application kernels can be run on any device.

```
std::vector<queue> getAllQueues(){
    auto devs = device::get_devices(info::
        device_type::gpu);
    std::vector<queue> queues(devs.size());
    for (int d=0; d < devs.size(); d++){
        queues[d] = queue(devs[d]);
    }
    return queues;
}
```

Figure 3: Example of a function for getting a vector of SYCL queues that attaches all the GPUs of the node.

In contrast, targeting multiple GPUs from different vendors using CUDA or HIP requires compiling each kernel native implementation utilizing the specific compiler and develop a host code capable of supporting memory management, data transfers and kernel launching. The host code is responsible for calling the right compiled version of the code, depending on the targeted platform. This imposes a significant extra development effort, compared to that necessary with Open SYCL.

However, to distribute the computation of one kernel across all devices, and to run all the sub-kernels concurrently, it is required that there are no data dependencies between sub-kernels; i.e., the range of the output data of each sub-kernel does not overlap any other sub-kernels' range. Otherwise, the SYCL runtime would serialize the execution of the sub-kernels after detecting the data dependencies, giving no advantage for using multiple GPUs. For example, let's suppose that the output of our kernel is an array of 1 000 elements, and we have two GPUs to execute the kernel. A non-overlapping distribution of the data could be the range [0, 511] for the first GPU and [512, 999] for the second, and the sub-kernels can run concurrently. An overlapping distribution of the data could be the range [0, 511] for the first GPU, and [500, 999] for the second; in this case, the execution of the sub-kernels would be serialized.

SYCL standard offers two ways to separate the ranges of the data: Ranged accessors and sub-buffers. A ranged accessor is an accessor constructed from a sub-range of a buffer, limiting the elements of the buffer that can be accessed. However, according to the SYCL standard, the ranged accessor creates a requisite for the entire buffer [53]. Therefore, since all the sub-kernels write the same buffer, their execution will be serialized, although each sub-kernel writes a non-overlapping range of the buffer. Regarding of the sub-buffers, they are buffers created from a sub-range of a buffer previously created. If two sub-buffers, $B_1$ and $B_2$, are created from the same buffer, but their ranges do not overlap, the accessors created from them, $A_1$ and $A_2$, will not overlap. Therefore, if a kernel $K_1$ uses $A_1$ and a kernel $K_2$ uses $A_2$, both kernels can be concurrently executed.

Unfortunately, Open SYCL does not currently support the sub-buffer feature. The only solution is to create a separated buffer for each sub-kernel, ensuring that the buffers ranges do not overlap. However, the compiler does not allow creating a vector of buffers, as other SYCL objects like queues. Moreover, the buffer cannot be created inside a *for* loop. Since each loop iteration creates a new scope, the SYCL runtime will create and destroy the buffer each iteration, and it will serialize the kernel execution instead of concurrently executing them.

Therefore, it is necessary to create one buffer for each possible sub-kernel, although the final number of executed sub-kernels is smaller. To illustrate this, Figure 4 shows an example of how the data is partitioned, assuming that there are three GPUs in the node (therefore, creating three buffers), but afterward using only two GPUs. At first, two vectors are created to store the offsets and ranges, being the vector size the maximum number of devices (lines 13 and 14). After that, the values of the vector are initialized. When the device *d* is used, the offset and range are calculated such that the data among sub-kernels is equally distributed (lines 17-20). If the device *d* is not used, we must also initialize the offset and range (lines 21-25).

After that, we should create the three buffers using the previously calculated offsets and ranges (lines 31-33). Note that,

```
1  [...]
2  //host memory declaration
3  int elements = 100000;
4  float host_array [elements];
5  //host memory inicialization
6  [...]
7  //Get all the possible queues
8  //Let's assume that there are 3 queues
9  auto queues = getAllQueues();
10 int numMaxDevices =  queues.size();
11 int usedDevices = 2;
12 //Create the offset and range vectors
13 std::vector<int> offset(numMaxDevices);
14 std::vector<int> ranges(numMaxDevices);
15 int chunk = elements / usedDevices;
16 for(int d=0; d<numMaxDevices; d++){
17     if(d < usedDevices){
18     //Used buffer, calculate range and offset
19         off[d] = chunk*d;
20         ranges[d] = chunk;
21     }else{
22     /*For unused buffers, ensure that the
           buffer can be created. This will not
           affect to the launched kernels, since
           the buffer is not used*/
23         off[d] = 0;
24         ranges[d] = 1;
25     }
26 }
27 /*Note that the last device will be compute the
       padded elements*/
28 ranges[usedDevices - 1] += elements %
       usedDevices;
29 //Start a new scope and create the buffers
30 {
31 buffer<float, 1> dev_buf0(host_array + off[0],
       range<1>{static_cast<size_t>(ranges[0])});
32 buffer<float, 1> dev_buf1(host_array + off[1],
       range<1>{static_cast<size_t>(ranges[1])});
33 buffer<float, 1> dev_buf2(host_array + off[2],
       range<1>{static_cast<size_t>(ranges[2])});
34 //launch the kernels
35 for(int i=0; i < usedDevices; i++)
36 {
37 //Create a pointer to the desired buffer
38    buffer<float,1> *usedBuf = (i==0) ? &dev_buf0
       : ((i==1)  ? &dev_buf1 :  &dev_buf2);
39    queues[i].submit([&](handler &my_handler){
40 //Create the accessor using the buffer pointer
41       auto array = usedBuf->get_access<
          access_mode::read_write>(my_handler);
42       //Launch the kernel
43       my_handler.parallel_for<class my_kernel>(
44       [...]
45       }); //end submit
46    }
47 }
48 } //end of scope and start the host code
49 [...]
```

Figure 4: Distributing kernel work on multiple GPUs using Open SYCL

although the third device is not used, the third buffer is always created (line 33). If this buffer is not correctly created, the application will be aborted when the invalid buffer is created. Correctly initializing the buffers ensures that the application works for a maximum of three devices, independently of the final number of used devices. In the example of Figure 4, the ranges of *dev_buf0*, *dev_buf1* and *dev_buf2* are [0, 49 999], [50 000, 99 999] and [0, 0], respectively. Note that, although the ranges of *dev_buf0* and *dev_buf2* are overlapped, this fact does not affect the concurrent execution of the two sub-kernels, since *dev_buf2* is never used and does not create data dependencies in the SYCL runtime.

Finally, the code starts a *for* loop with *usedDevices* iterations (line 35). At each iteration, a pointer to the appropriate buffer is created, called *usedBuf* (line 38). Then, the kernel is submitted to the queue *i*, and *usedBuf* is used to create the accessor (line 41) that will be used inside the kernel.

Using the buffers this way allows distributing the computation between several GPUs, but it increases the development effort of the code, as it will be seen in Section 6. Note that the example of Figure 4 only works for a maximum of three GPUs. If the target system has six GPUs, it is necessary to add three more buffers and modify the buffer selection when *usedBuf* is created. This extra development effort is greater when the number of GPUs or the number of data structures to distribute increases. This does not happen with the native versions, which can run with any number of GPUs without modifications. However, combining NVIDIA and AMD GPUs is easier using SYCL than combining the CUDA and HIP native versions, as explained at the beginning of this section.

Finally, another consideration that should be taken into account is that, although Open SYCL has support for simultaneously executing kernels in NVIDIA and AMD GPUs, it has no support for transparently perform data transfers between both architectures. This can be solved in two ways: 1) Manually transferring data from one device to another through the host, or 2) ensuring that there are no data dependencies between the devices of the different vendors. In our case, the second one is the best option, since the data has already been distributed avoiding data dependencies, thus ensuring the concurrent execution of all the sub-kernels.

Summarizing, the steps to enable using multiple GPUs in the Open SYCL version of the UVaFTLE, assuming that our system has four GPUs, are the following:

1. Get a vector of *queues* to allow using all the GPUs.
2. Calculate the range and offset of each *sub-kernel* for:
   (a) The output array of preprocessing kernel (also used as an input in the second kernel).
   (b) The output array of FTLE kernel.
3. Start a new scope and define four *buffers* to manage the output array of the preprocessing kernel, using the ranges and offset previously calculated.
4. Define four *buffers* to manage the output array of the FTLE kernel.
5. Start a *for* loop with one iteration per used device. At each iteration:
   (a) Create a buffer pointer (*p_preproc*), associated to the appropriate output buffer of the preprocessing kernel.
   (b) Create a buffer pointer (*p_ftle*), associated to the appropriate output buffer of the FTLE kernel.
   (c) Submit the preprocessing kernel, create the output accessor from *p_preproc*, and launch it using an *nd-range parallel for*.
   (d) Submit the FTLE kernel, create an input accessor from *p_preproc*, the output accessor from *p_ftle*, and launch it using a *nd-range parallel for*.
6. End the *for* loop and the scope to update the host memory, allowing the host to get the final results.

## 5. Performance evaluation

In this section, we first describe the platform where the experiments have been conducted. Then we list the test cases, and afterward we summarize the execution times observed when targeting AMD GPUs, NVIDIA GPUs, and a combination of them.

### 5.1. Platform

The experiments have been conducted in a computing server property of the *Universidad de Valladolid* which features two Intel(R) Xeon(R) Platinum 8160 CPU @ 2.10GHz, with 24 Core Processors and 48 physical threads each, an NVIDIA Tesla V100 PCIe 32 GB GPU, and an AMD Vega 10 XT Radeon PRO WX 9100 GPU. The server is equipped with a CentOS 7 operating system. The toolchains used are GCC 11.1, CUDA 11.3, ROCm 5.4.3, and LLVM 14.0.6. This LLVM distribution has been used to compile the Open SYCL compiler, whose version is 0.9.4.

### 5.2. Test cases

To conduct the performance evaluation, we have chosen two applications widely used in the literature when evaluating flowmap and FTLE computations: The Double-Gyre flow [54] for the 2D case, and the Arnold–Beltrami–Childress (ABC) flow or Gromeka–Arnold–Beltrami–Childress (GABC) flow [55] for the 3D case. In particular, our evaluation in the 2D case uses a mesh composed of 10 000 000 points, and in the 3D case a mesh composed of 1 000 000 points. Table 2 reflects the details associated to each mesh geometry: The dimensions, the number of mesh points and mesh simplex (either triangles or tetrahedrons), the interval of interest at each axis, and the number of elements in the interval at each axis taken to define the mesh points.

|  | 2D | 3D |
|---|---|---|
| **Dim** | ≈10 000K (9 998 244) | 1 000K |
| **nFaces** | 19 983 842 | 5 821 794 |
| **min-max(x, y, z)** | (0-2, 0-1, 0-0) | (0-1, 0-1, 0-1) |
| **length(x, y, z)** | (3 162, 3 162, 0) | (100, 100, 100) |

Table 2: Description of the test cases used in our experiments.

For each described FTLE test case, we evaluate the performance (in terms of execution time) using both AMD and NVIDIA architectures with different compiler options, as follows:

- **Native code, native compiler**: The CUDA/HIP code has been compiled using the vendor toolchain (nvcc/hipcc).

- **Native code, clang compiler**: The CUDA/HIP code has been compiled using clang compiler included in the LLVM toolchain.

- **SYCL code, Open SYCL compiler**: The SYCL code has been compiled using Open SYCL compiler, using the SCMP model (see Section 3). This model also allows testing the code using the four GPUs of the system, using simultaneously AMD and NVIDIA GPUs.

Note that, using Open SYCL with the SCMP model, the generation of the architecture code relies on the clang compiler, not present in Open SYCL [17]. For this reason, we have compiled the native code using the same LLVM toolchain used to compile Open SYCL. The vendor compilers have been also used to compile the FTLE application. However, since the code generation is not made by the same tool, performance comparison results with SYCL are unexpected.

Each test has been repeated 30 times and the results shown reflect the average of all of them. Note that, when a kernel is executed using two or more GPUs, we take the longest execution time observed for all the sub-kernels, this is, the one associated to the slowest sub-kernel execution. Moreover, all the results we will show in the following sections are those associated to the executions using pinned memory for memory transfers, as those are always slightly better than the ones observed without pinned memory.

Finally, we want to highlight that the preprocessing kernel takes more time to be executed than the FTLE kernel. Thus, the execution time shown for the first kernel is reflected in seconds and, for the second one, in milliseconds.

*5.3. Performance results targeting AMD GPUs*

In this section, we summarize the results observed when targeting AMD GPU devices. Figure 5 illustrates the results observed for the 2D and 3D FTLE test cases, detailing the execution time observed for the preprocessing and FTLE kernels when using, in each case, 1 or 2 GPUs with the HIP, CLANG and SYCL compilers. Note that *HIP (hipcc)* refers to the HIP-based version of the UVaFTLE compiled using the hipcc compiler, *HIP (clang)* refers to the HIP-based version compiled using the clang compiler, and *SYCL* is the SYCL-based version compiled using the Open SYCL compiler.

Based on the presented results, the first observation is that, in all cases, using a second GPU reduces the execution time with respect to the use of a single GPU, because the computational load is spread among the devices. However, we observe proportionally better results for the FTLE kernel than for the preprocessing kernel. This is due to the fact that the preprocessing kernel is memory-intensive, performs numerous global memory accesses, and, additionally, the number of elements accessed by each thread is not homogeneous; contrarily, the FTLE kernel dedicates the majority of its time to solve arithmetic operations. We observe that the mentioned trend is applicable both for the 2D and 3D test cases. Regarding the preprocessing kernel, the performance differences observed when using different programming models and compilers are too small to be considered significant.

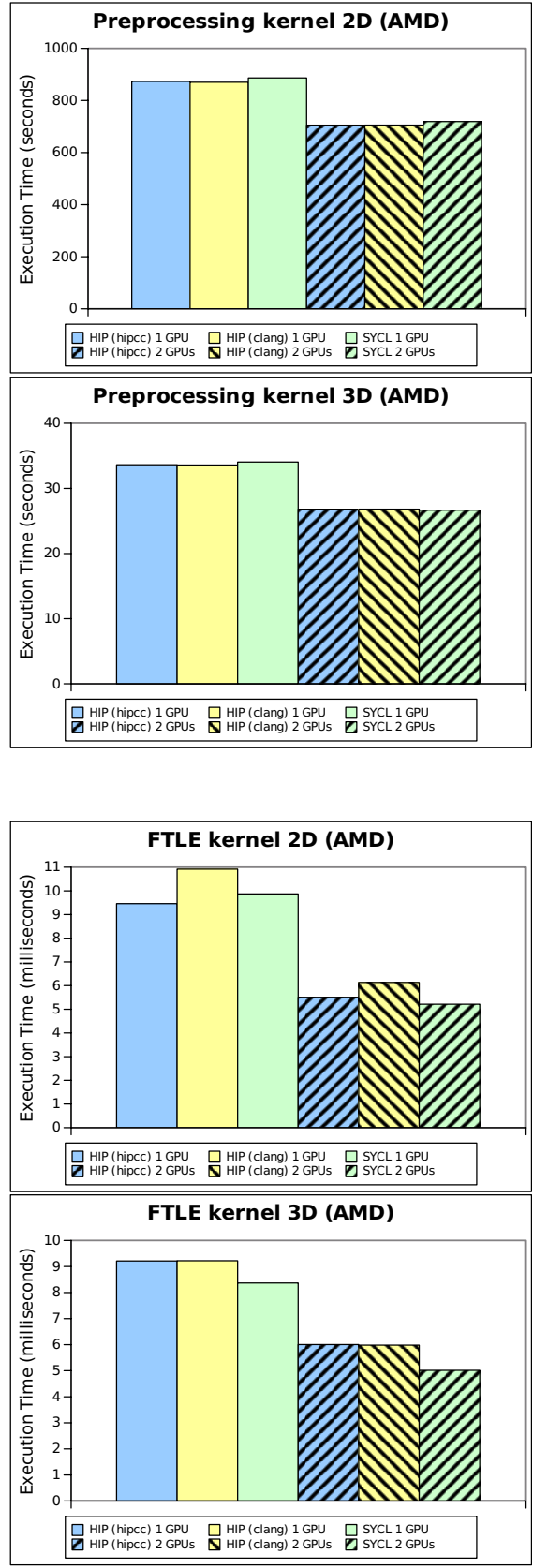

Figure 5: Performance evaluation results of each kernel when targeting AMD GPU devices for the 2D and 3D FTLE test cases.

*5.4. Performance results targeting NVIDIA GPUs*

In this section, we summarize the results observed when targeting NVIDIA GPU devices. Figure 6 illustrates the results observed for the 2D and 3D FTLE test cases, detailing the execution time observed for the preprocessing and FTLE kernels when using, in each case, one or two GPUs with the CUDA, CLANG and SYCL compilers. In this case, *CUDA (nvcc)* refers to the CUDA-based version of the UVaFTLE compiled using the nvcc compiler, *CUDA (clang)* refers to the CUDA-based version compiled using the clang compiler, and *SYCL* is the SYCL-based version compiled using the Open SYCL compiler.

As stated in the previous section, here we again observe that, in all cases, using a second GPU reduces the execution time. For the same reason formerly detailed, we also observe better results for the FTLE kernel than for the preprocessing kernel.

In contrast to what we observed with AMD GPU devices, here there are remarkable differences when comparing the results offered by the different programming models and compilers. In the preprocessing kernel, the CUDA (nvcc) version is always the fastest one, and the other two do not show remarkable differences. Nevertheless, in the case of the FTLE kernel, the results observed can be considered equivalent in any case, as the differences are smaller than 0.2 milliseconds.

Finally, we observe that the mentioned trends are applicable both for the 2D and 3D test cases.

*5.5. Performance results targeting NVIDIA and AMD GPUs*

In this section, we summarize the results observed when targeting AMD and NVIDIA GPU devices simultaneously. Figure 7 illustrates the results observed for the 2D and 3D FTLE test cases detailing the execution time observed for the preprocessing and FTLE kernels when using, in each case, one AMD or NVIDIA GPU, two GPUs, either AMD or NVIDIA, and four GPUs (two of each vendor) using the Open SYCL compiler.

The first thing we observe is that NVIDIA execution time is much smaller than that provided by AMD devices. This is due to the higher computational power and peak performance of the available devices of each vendor in our system.

Secondly, in the FTLE kernel case, we observe that it is better to use two NVIDIA GPU devices than the four GPUs. This difference is due to one NVIDIA GPU executes the FTLE kernel four times faster than one AMD GPU. Therefore, although the computation is divided into four equal parts, the AMD GPU requires the same time to compute a quarter of the calculations as one NVIDIA GPU to do the whole calculation. Nevertheless, this does not happen with the preprocessing kernel, where the best performance is obtained when the four GPU devices are simultaneously used. As commented in Section 5.3, the load of each preprocessing sub-kernel is not homogeneous, therefore, by submitting the lightly loaded sub-kernels to the AMD GPUs, and the heavily loaded sub-kernels to the NVIDIA GPUs, we can accelerate the whole kernel execution.

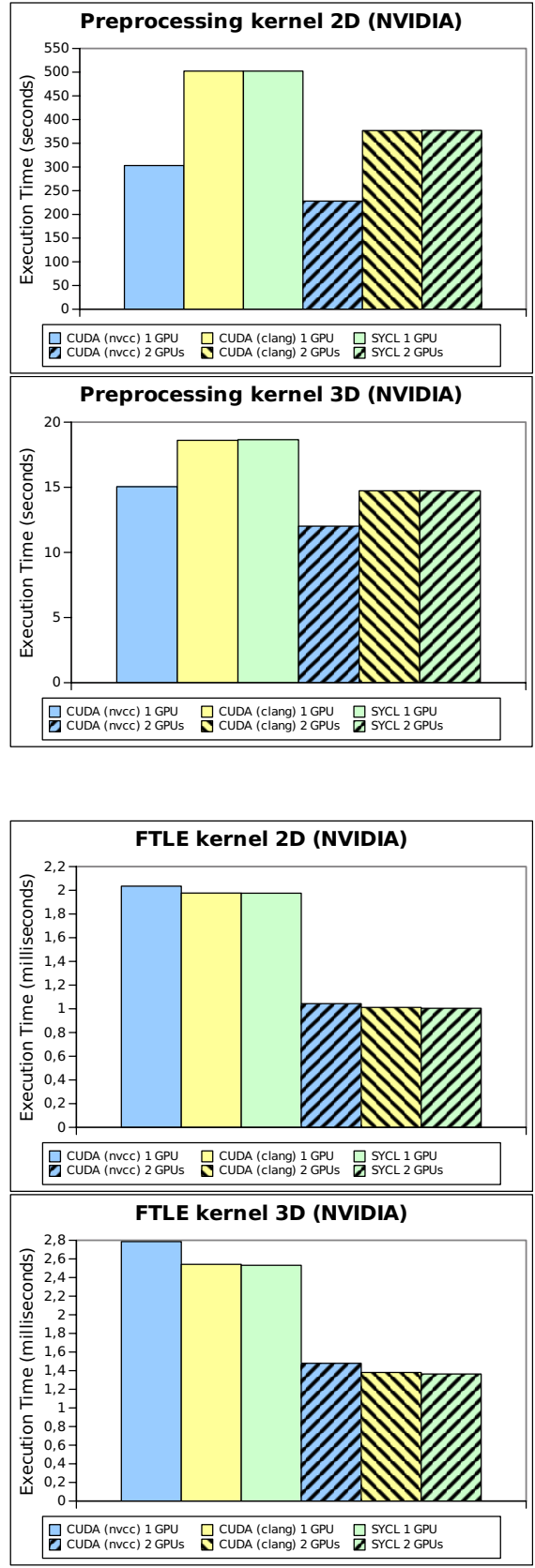

Figure 6: Performance evaluation results of each kernel when targeting NVIDIA GPU devices for the 2D and 3D FTLE test cases.

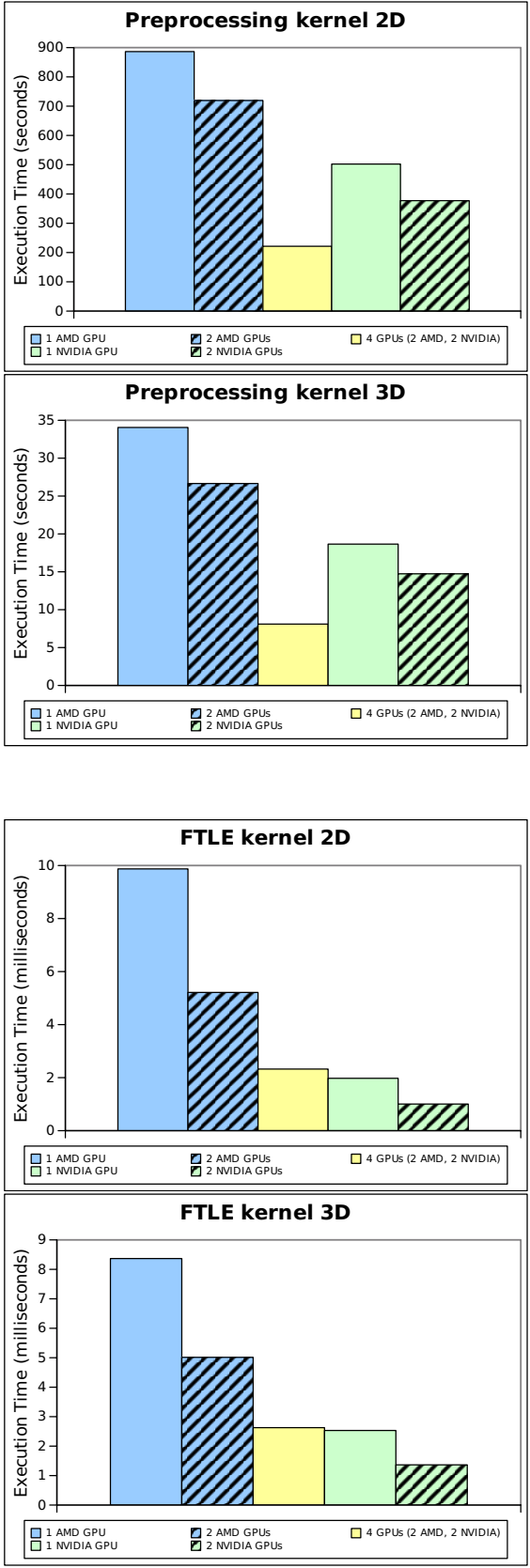

Figure 7: Open SYCL performance evaluation results of each kernel targeting AMD and NVIDIA GPUs simultaneously for the 2D and 3D FTLE test cases.

Finally, we want to highlight the most important conclusion of this section experiment: Open SYCL allows us to exploit multiple GPUs from different vendors simultaneously. This allows the exploitation of more heterogeneous clusters where different devices with different computational powers can be exploited by creating a better load-balance for both the sub-kernels in the preprocessing kernel, and for any stage of the computation in future versions of the program.

## 6. Development effort

In this section, we analyze the differences in development effort between CUDA, HIP and SYCL codes of UVaFTLE. We consider four classical development effort metrics: The number of lines of code (LOC), the number of code tokens (TOK), McCabe's cyclomatic complexity (CCN) [56], and Halstead's development effort [57]. The first two metrics measure the code volume that the user should program. The third measures the rational effort required to program it in terms of code divergences and potential issues that should be considered to develop, test, and debug the program. The last metric measures both code complexity and volume indicators, obtaining a comprehensive measure of the development effort.

The measured codes include the data structures management, the kernel definitions, and the coordination host code. For a fair comparison, each version is written in a single source-code file, and all versions have been formatted following the same criteria. The differences between codes are the strictly necessary ones, associated to the particularities of each programming model. For example, comparing the FTLE kernels in CUDA and SYCL, the main differences are how the thread global index is calculated, as explained in Section 4.4, and certain calls to perform mathematical operations, such as square root or cosine. The CUDA and HIP versions of the program support multiple GPUs of the corresponding vendor. As we explain in Section 4.5, by enabling multi-GPU execution, the final SYCL code changes in volume depending on the maximum number of GPUs allowed. For this reason, we have compared four versions of the SYCL code, allowing a maximum of 1, 2, 4, and 8 GPUs, respectively. The cleaned versions of both the SYCL programs and the CUDA and HIP versions can be found in our repository, in the folder *measure-codes* [13].

Table 3 reflects the measures of the four development-effort metrics for each one of the functions that present changes that depend on the programming model chosen. They include the three critical functions that have been transformed into kernels (preprocessing, and the 2D and 3D FTLE functions), and the main function, that contains the memory management and kernel calls. Table 4 reflects the measures of the four development effort metrics considering the whole program, which includes the functions and kernels reflected in Table 3, and other auxiliary functions and declarations that do not depend on the heterogeneous programming model selected.

The metrics reveal that the development effort of CUDA and HIP versions are almost the same. Their kernels are identical,

| Function/Kernel | Code version | LOC | TOK | CCN | Halstead |
|---|---|---|---|---|---|
| Preprocessing | CUDA | 19 | 190 | 8 | 23 908 |
| | HIP | 19 | 190 | 8 | 23 908 |
| | SYCL | 25 | 361 | 9 | 71 779 |
| FTLE 2D | CUDA | 134 | 1090 | 26 | 508 649 |
| | HIP | 134 | 1090 | 26 | 508 649 |
| | SYCL | 144 | 1338 | 27 | 676 094 |
| FTLE 3D | CUDA | 194 | 1785 | 40 | 918 499 |
| | HIP | 194 | 1785 | 40 | 918 499 |
| | SYCL | 204 | 2097 | 41 | 1 228 892 |
| main | CUDA | 196 | 1657 | 17 | 650 334 |
| | HIP | 196 | 1644 | 17 | 614 989 |
| | SYCL 1 GPU | 167 | 1544 | 19 | 636 370 |
| | SYCL 2 GPUs | 171 | 1651 | 21 | 696 478 |
| | SYCL 4 GPUs | 175 | 1823 | 25 | 804 617 |
| | SYCL 8 GPUs | 183 | 2211 | 33 | 1 076 951 |

Table 3: Development effort metrics for each function/kernel, according to the programming model employed.

| Code Version | LOC | TOK | CNN | Halstead |
|---|---|---|---|---|
| CUDA | 645 | 5302 | 110 | 5 315 886 |
| HIP | 643 | 5289 | 110 | 5 282 624 |
| SYCL 1 GPU | 645 | 5976 | 116 | 8 328 659 |
| SYCL 2 GPUs | 649 | 6083 | 118 | 8 473 727 |
| SYCL 4 GPUs | 653 | 6255 | 122 | 8 771 842 |
| SYCL 8 GPUs | 661 | 6643 | 130 | 9 496 005 |

Table 4: Development effort metrics for the whole code, according to the programming model employed.

and the differences in the main code are almost negligible in terms of LOC, TOK, and CCN, and very small considering the Halstead results (a little mode than 1% higher in the case of CUDA).

Regarding the SYCL version of the kernels (Table 3), the values measured for the four metrics are higher compared to the CUDA/HIP results. Nevertheless, the CCN results present almost the same values as those observed for the native versions. This LOC and TOK higher values are mainly due to the *submit* lambda function, the *nd-range parallel for* lambda function, and the creation of the *accessors*. The preprocessing kernel is the most affected one by this increase, as it is the smallest kernel, being its code lines increased in 31% and its number of tokens in 90%. The Hasltead's development effort is three times higher in SYCL than in the other two versions.

This difference is less significant in the other two kernels: 7% more lines, 22% more tokens and 32% mode Halstead's development effort for the 2D kernel, and 5% more lines, 17% more tokens and 34% more Halstead's development effort for the 3D kernel. These measures indicate that the increase of development effort is greater with small kernels than with large kernels due to the minimum programming structures, declarations, and initialization needed in a SYCL kernel.

Analyzing the main function of the code (Table 3), we observe that the SYCL version for one GPU has fewer LOC, TOK and Halstead's measures than the native versions, thanks to the transparent memory management through the *buffer* abstraction. However, these metrics increase as the maximum number of allowed GPUs increases, because of the multiple SYCL queues that are necessary, as explained in Section 4.5. Even so, the LOC measure never exceeds the native version, although the TOK is practically the same for 2 GPUs, and it is greater for 4 or 8 GPUs. Halstead's measures for the versions with more than one GPU are always higher than for the vendor specific models. In contrast, the CCN is always greater in the SYCL version, and it increases with the number of GPUs supported due to the extra logic for the management of the different queues.

Finally, if we analyze the whole code (Table 4), it can be seen that the SYCL code has greater development effort metrics than native versions, even in single GPU version, and specially for the TOK and Halstead metrics. The only exception to that is the LOC value when using a single GPU with SYCL, which is exactly the same as that corresponding to CUDA.

As summary, we observe that in SYCL the transparent man-agement of buffers and memory movements for a single device and queue are simpler than orchestrating the equivalent asynchronous operations in CUDA or HIP. However, the elaborated syntax and declarations needed for kernels increase their complexity, specially for simple or small kernels. Moreover, in the SYCL host code, the management of each extra device introduces more complexity, while in the CUDA and HIP versions the management of an arbitrary number of devices can be easily abstracted. However, this SYCL problem could be solved in the future if the compilers include full support for sub-buffers (see Section 4.5).

## 7. Concluding remarks

There are several proposals for high-level heterogeneous programming that try to reduce the development effort while improving functional and performance portability. SYCL is one of the proposals with a higher impact of the community, due to the abstractions proposed and the evolution of its compilers and programming frameworks, which are reaching a higher maturity level. This paper evaluates the SYCL programming model, using the Open SYCL compiler, from two different perspectives: (1) The performance it offers when dealing with single or multiple GPU devices of the same or different vendors; and (2) the development effort required to implement the code. For this purpose, we use as case of study the FTLE application over two real-world scenarios: The Double-Gyre flow (2D) and the ABC flow (3D). The evaluation is based on a comparison of a SYCL implementation vs. baseline codifications using the specific programming tools for two GPU vendors: CUDA for NVIDIA GPUs and HIP for AMD GPUs.

The main conclusions that can be extracted from this work are:

- The performance results reveal that there is not a remarkable overhead associated to SYCL usage in terms of the GPU kernel execution times, compared to the performance obtained when using kernel native implementations based on CUDA or HIP. The only case when this is not true is when comparing the FTLE kernel CUDA based version compiled with nvcc against that same one compiled with clang, or the equivalent Open SYCL version, as the first one is clearly faster.

- We have evaluated two kernels that are very different in terms of their nature: As explained in previous sections, the preprocessing kernel is much more memory intense than the FTLE one, which focuses on solving a collection of linear algebra operations and is much faster to be completed. Thanks to comparing the performance using both of them, we can affirm that the scalability observed with the native versions and Open SCYL is equivalent, although the kernels' typology is very different.

- Regarding the multi-GPU executions with Open SYCL when using four GPU devices, being 2 from NVIDIA and 2 from AMD, it is important to first highlight that the code is able to leverage simultaneously all of them. Moreover, the performance results observed reflect that using the four GPU devices improves the results for the preprocessing kernel. However, this is not true for the FTLE kernel, because it submits non-homogeneously loaded sub-kernels that suffer from the computational power difference between the available AMD and the NVIDIA devices in our system.

- The development effort measures indicate that, in SYCL, the transparent management of buffers and memory movements for a single device and queue are simpler than programming asynchronous operations in CUDA or HIP. However, the basic kernel syntax and the declarations needed are more complex in SYCL, which is more noticeable in small or simple kernels.

- With the current development status of the Open SYCL compiler, the development effort metrics reveal that the management of each extra device introduces more code complexity, while in the CUDA and HIP versions the management of an arbitrary number of devices can be easily abstracted. Nevertheless, although the development effort increases, the SYCL programs are more portable, and can run the application distributing the computation in both NVIDIA and AMD GPUs, even combining GPUs of the two vendors in the same execution. With vendor provided models, this could be done by combining them in a much more complicated code that should include the solutions in both models, and adding some kind of data communication across them.

As part of the future work, we plan to explore how a better load balancing in the preprocessing kernel affects SYCL performance, compared to CUDA and HIP implementations. Moreover, we also plan to explore the usage of other SYCL implementations/compilers to target alternative computational devices, such as FPGAs, to conduct a similar evaluation.

## Acknowledgment

This work was supported in part by the *Spanish Ministerio de Ciencia e Innovación* and by *the European Regional Development Fund (ERDF)* program of the European Union, under Grant PID2022-142292NB-I00 (NATASHA Project); and in part by *the Junta de Castilla y León - FEDER Grants*, under Grant VA226P20 (PROPHET-2 Project), Junta de Castilla y León, Spain. This work was also supported in part by grant TED2021–130367B–I00, funded by *European Union NextGenerationEU/PRTR* and by *MCIN/AEI/10.13039/501100011033*. This work has been also partially supported by NVIDIA Academic Hardware Grant Program.